\documentclass[journal]{IEEEtran}
\usepackage{amsmath,amsfonts}
\usepackage{algorithmic}
\usepackage{algorithm}
\usepackage{lipsum}
\usepackage{array}
\usepackage[caption=false,font=normalsize,labelfont=sf,textfont=sf]{subfig}
\usepackage{textcomp}
\usepackage{stfloats}
\usepackage{url}
\usepackage{verbatim}
\usepackage{graphicx}
\usepackage{cite}
\usepackage{color}
\usepackage{xcolor}
\usepackage{hyperref}
\usepackage{ragged2e}
\usepackage{stfloats} 
\usepackage{autobreak}
\usepackage{cuted}
\usepackage{authblk}
\usepackage{amssymb}

\begin{document}
\title{Collaborative Intelligence for UAV-Satellite Network Slicing: Towards a Joint QoS-Energy-Fairness MADRL Optimization}
\author[1]{Thanh-Dao Nguyen}
\author[1]{Ngoc-Tan Nguyen}
\author[1]{Thai-Duong Nguyen}
\author[2]{Nguyen Van Huynh}
\author[3]{Dinh-Hieu Tran}
\author[3]{Symeon Chatzinotas}
\affil[1]{\textit{University of Engineering and Technology, Vietnam National University, Hanoi, Vietnam}}
\affil[2]{\textit{School of Computer Science and Informatics, University of Liverpool, United Kingdom}}
\affil[3]{\textit{Interdisciplinary Centre for Security Reliability and Trust (SnT), University of Luxembourg, Luxembourg}}
\maketitle
\begin{abstract}
Non-terrestrial networks are critical for achieving global 6G coverage, yet efficient resource management in aerial and space environments remains challenging due to limited onboard power and dynamic operational conditions. Network slicing offers a promising solution for spectrum optimization in UAV-based systems serving heterogeneous service demands. For that, this paper proposes a hierarchical network slicing framework for UAV–satellite–integrated networks supporting eMBB, URLLC, and mMTC services. Specifically, we formulate a joint optimization of UAV trajectory, transmission power, and spectrum allocation as a decentralized partially observable Markov decision process that ensures quality of service while minimizing energy consumption and maximizing resource fairness. To address the computational intractability and partial observability, we develop a multi-agent deep reinforcement learning solution under the centralized training and decentralized execution paradigm. In the proposed system, UAV agents act as distributed actors coordinated by a shared critic operating with multi-head attention mechanism at a low Earth orbit satellite. Experimental results then demonstrate that our approach outperforms existing methods by up to 33\% in cumulative reward while achieving superior energy efficiency and fairness.
\end{abstract}



\section{Introduction}

6G networks must be able to serve a wide variety of smart systems, each with heterogeneous and stringent service requirements \cite{jiang2021roadtowards6G}. However, terrestrial infrastructure alone cannot provide sufficient support due to geographical constraints, deployment costs, and vulnerability to disasters. Non-terrestrial networks, particularly UAV-enabled systems, address these limitations through rapid deployment and flexible repositioning\cite{azari2022EvoOfNTN}. Despite advantages, UAVs face critical challenges including limited onboard energy, constrained spectrum resources, and the need for adaptation to dynamic environments.

Network slicing emerges as an enabling technology to address these shortcomings by virtualizing shared physical resources into isolated logical networks tailored to specific service requirements \cite{bouzid2023taxanomyNS, mishra2025HetNet}. By integrating UAV mobility with network slicing, operators can simultaneously serve  users of different service types from a common aerial infrastructure. This integration is particularly valuable in scenarios where terrestrial networks are unavailable, such as disaster recovery, maritime operations, or temporary mass gatherings, while optimizing the limited energy and spectrum budgets of UAVs.

Recent research on UAV network slicing can be broadly categorized into optimization-based and learning-based approaches. Optimization methods \cite{wei2024hierarchicalNS, yang2021proactiveUAV-NS} typically model the resource allocation problem as mixed-integer programs or convex approximations, providing optimal solutions under simplified assumptions. However, these methods suffer from high computational complexity and limited adaptability to dynamic environments. On the other hand, machine learning approaches \cite{haiyan2024PriorityLoadBalancing, bellone2023DRLforCoverageandResourceallocation, cui2020MARLforRA, carrillo2022optimizeconnectivity} leverage deep reinforcement learning (DRL) to learn adaptive policies through interaction with the environment, offering better scalability and real-time performance. Despite these advances, three critical gaps remain. First, most existing works \cite{wei2024hierarchicalNS, yang2021proactiveUAV-NS, bellone2023DRLforCoverageandResourceallocation, cui2020MARLforRA, cui2024vehicularNet} assume UAVs operate as supplements to terrestrial base stations. This limits their applicability in scenarios where terrestrial infrastructure is not available. Second, while full 3D trajectory control is essential for coverage and interference optimization, many studies either ignore UAV trajectory optimization entirely \cite{haiyan2024PriorityLoadBalancing, carrillo2022optimizeconnectivity}, consider only vertical movement \cite{wei2024hierarchicalNS}, or assume fixed altitudes \cite{bellone2023DRLforCoverageandResourceallocation}. Finally, existing DRL frameworks typically optimize a single objective (e.g., throughput) or treat multiple objectives separately. A unified framework that jointly optimizes trajectory, power control, and bandwidth allocation while balancing QoS, energy efficiency, and fairness remains absent.

To address these limitations, this paper makes three key contributions. First, we develop a UAV  network architecture integrated with LEO satellite systems, enabling resilient connectivity across diverse scenarios such as natural disasters, remote regions, and temporary high-demand events where terrestrial infrastructure is inadequate. Second, we propose using a multi-agent deep deterministic policy gradient framework with a shared critic utilizing multi-head attention mechanism. This framework enables UAV agents to cooperatively learn optimal trajectories, transmission power control, and spectrum allocation strategies while maintaining distributed operation during execution. Finally, through extensive simulations, we demonstrate that the proposed framework allows agents to exhibit emergent cooperative behaviors that jointly optimize QoS satisfaction, energy efficiency, and user fairness, outperforming baseline methods by up to 33\% in cumulative rewards.

\vspace{-0.2em}

\section{System Model}
\vspace{-0.2em}
\begin{figure}[ht]
    \centering
    \includegraphics[width=1\linewidth]{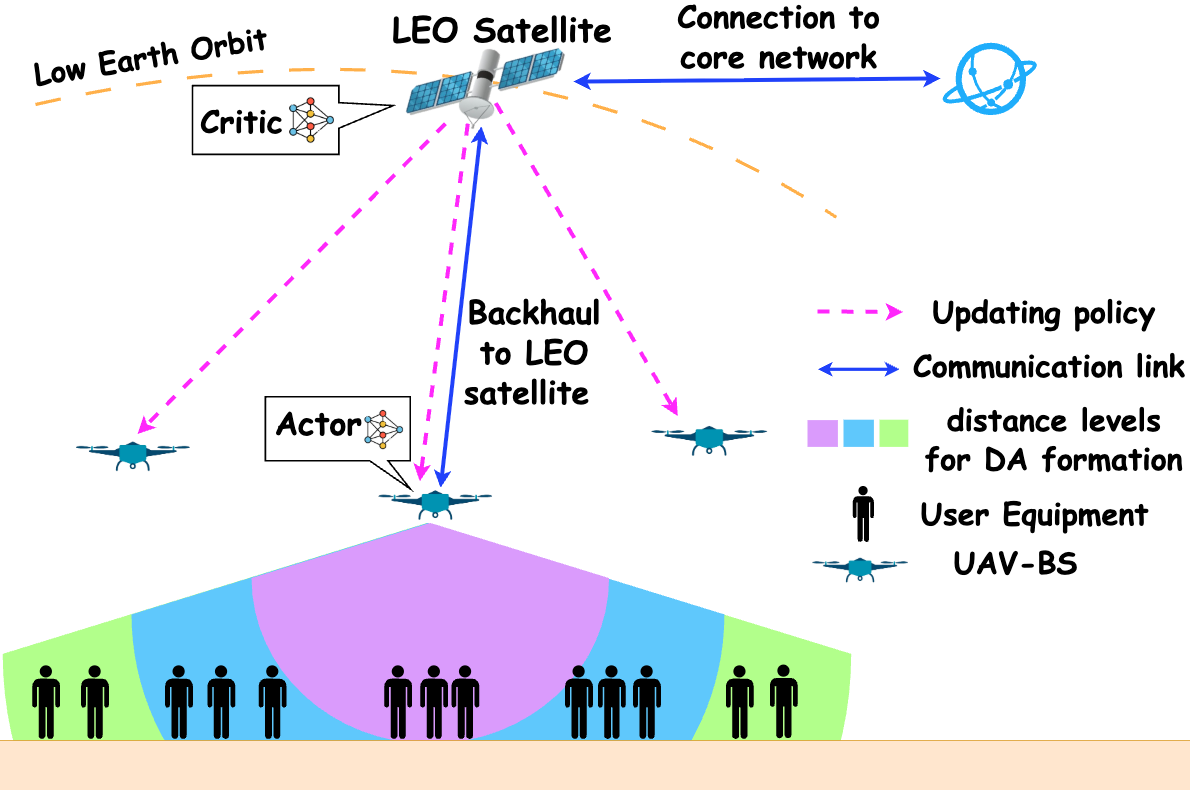}
    \caption{System model of standalone non-terrestrial UAV-Satellite network.}
    \label{fig:sys-mod}
\end{figure}

We consider a standalone non-terrestrial network comprising UAV-based base stations (UAV-BSs) and LEO satellites, designed to provide downlink services to ground users. As illustrated in Fig.~\ref{fig:sys-mod}, UAV-BSs provide access services while a LEO satellite provides backhaul connectivity and hosts the centralized training infrastructure. The set of UAV-BS is defined as $\mathcal{U} = \{u_1, u_2, ..., u_{|\mathcal{U}|}\}$, where each UAV $u$ operates at $\mathbf{p}_u= (x_u, y_u, h_u)$ under spatial constraints $x_u \in [x_{\min}, x_{\max}]$, $y_u \in [y_{\min}, y_{\max}]$, and $h_u \in [h_{\min}, h_{\max}]$. Each UAV has a maximum bandwidth $B_u^{\max}$, and can adjust its transmit power $P_u$ for energy and interference management. Under a universal frequency reuse scheme, all UAVs share the same spectrum, introducing inter-UAV interference that must be mitigated. The system supports three logical slices $\mathcal{S} = \{1, 2, 3\}$ representing eMBB, URLLC, and mMTC, respectively. Each slice $s$ is characterized by a minimum throughput requirement $R_s$. The set of user equipments (UEs) are denoted by $\mathcal{E} = \{e_1, e_2, ..., e_{|\mathcal{E}|}\}$, each located at $(x_e, y_e, h_e)$ and associated with slice $s_e \in \mathcal{S}$.

Within this framework, UAV-BSs collaboratively optimize trajectory planning, transmission power, and spectrum allocation across network slices to enhance QoS, energy efficiency, and user fairness. To enable intelligent spectrum slicing and trajectory adaptation, each UAV-BS hosts a DRL agent that interacts with the environment to collect state–action–reward experiences. These experiences are aggregated at an edge–cloud server on a LEO satellite, which performs centralized training via a shared critic network to guide distributed UAV actors.

\vspace{-0.5em}

\subsection{Channel Model}

We adopt an air-to-ground path loss model \cite{alhouraniA2GPropModel}, where the path loss (dB) between UAV $u$ and UE $e$ is modeled as:
\begin{equation}
\text{PL}(d_{e,u}, \theta_{e,u}) = 10\log_{10}\left(\frac{4\pi d_{e,u}}{\lambda}\right)^2 + \eta(\theta_{e,u}),
\label{eq:path-loss}
\end{equation}
where $d_{e,u}$ is the Euclidean distance between UAV $u$ and UE $e$, $\lambda$ is the carrier wavelength, and the first term represents the free-space path loss. The excessive path loss $\eta(\theta_{e,u})$ captures the impacts of line-of-sight (LOS) and non-line-of-sight (NLoS) propagation conditions, where $\theta_{e,u}$ is the elevation angle from UE $e$ to UAV $u$. The excessive loss is computed as a weighted average of LoS and NLoS components:
\begin{equation}
\eta(\theta_{e,u}) = \eta_{\text{LoS}} \cdot P_{\text{LoS}}(\theta_{e,u}) + \eta_{\text{NLoS}} \cdot (1 - P_{\text{LoS}}(\theta_{e,u})),
\label{eq:excessive-loss}
\end{equation}
where the LoS probability follows the sigmoid model \cite{alhourani2014OptCoverage}: $P_{\text{LoS}}(\theta) = 1/(1 + a \exp(-b(\theta - a)))$. Parameters for suburban environment are used: $a = 4.88$, $b = 0.43$, $\eta_{\text{LoS}} = 0.1$ dB, and $\eta_{\text{NLoS}} = 21$ dB  \cite{alhourani2014OptCoverage}. The received signal power at UE $e$ from UAV $u$ is measured as: $P_{e,u}^{\text{rx}} = P_u \cdot 10^{-\text{PL}(d_{e,u}, \theta_{e,u})/10}$  W.

Under universal frequency reuse, UE $e$ experiences interference from all non-serving UAVs. The signal-to-interference-plus-noise ratio (SINR) at UE $e$ served by UAV $u$ is:
\begin{equation}
\text{SINR}_{e,u} = \frac{P_{e,u}^{\text{rx}}}{\sum_{u' \in \mathcal{U} \setminus \{u\}} P_{e,u'}^{\text{rx}} + N_0 B_{e}},
\label{eq:sinr}
\end{equation}
where $N_0 = 10^{-13}$ W/Hz is the noise power spectral density and $B_{e}$ is the bandwidth allocated to UE $e$ (in Hz). The achievable throughput for UE $e$ is then given by the Shannon capacity formula: $T_e = B_e \log_2(1 + \text{SINR}_{e,u})$. The total delay $D_e$ comprises: propagation ($D^{\text{prop}} = d_{e,u}/c$), transmission ($D^{\text{trans}} = L_e/T_e$), retransmission ($D^{\text{retx}}$ 4 transmission attempts), queuing ($D^{\text{queue}}$ following M/D/1 model), handover ($D^{\text{ho}} = 50$ ms), and processing ($D^{\text{proc}} = 3$ ms). Reliability is modeled as $R_e = (1 - \text{PER}_e)^4 (1 - p_{\text{drop}})$, where packet error rate $\text{PER}_e$ is derived from SINR using standard block error rate model, and $p_{\text{drop}}$ is the buffer overflow probability.

\vspace{-0.5em}
\subsection{Hierarchical Resource Allocation}

\begin{figure}[h]
    \centering
    \includegraphics[width=0.95\linewidth]{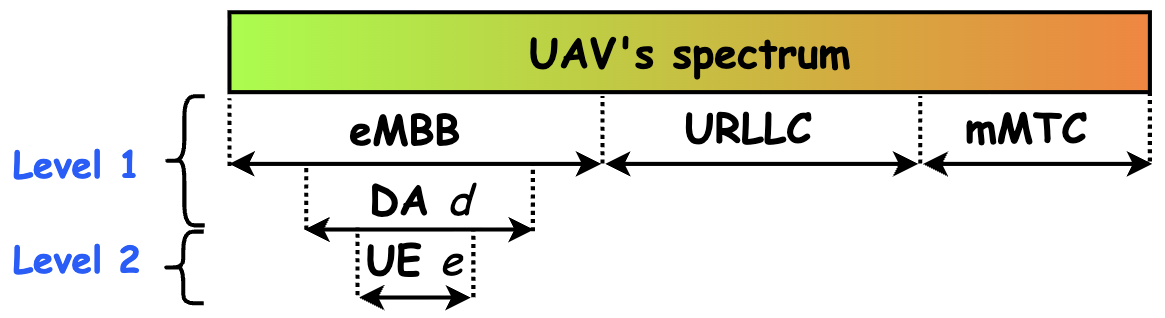}
    \caption{Two-level spectrum allocation framework}
    \label{fig:spec-division}
\end{figure}

Within this research, we adopt a hierarchical allocation framework using Demand Areas (DAs) as an intermediate abstraction layer \cite{wei2024hierarchicalNS}. Instead of allocating bandwidth directly to UEs, which would create an action space growing linearly with the number of UEs, we group users into DAs based on: (i) UEs' slice types (eMBB, URLLC, or mMTC), and (ii) Distance from UEs to the serving UAV (reflecting similar channel conditions). This yields a fixed number of DAs for each UAV (product of slice types and distance levels), enabling UAVs to make allocations at the DA level rather than per-user. 

As illustrated in Fig.~\ref{fig:spec-division}, resource allocation proceeds hierarchically. At level 1, each UAV distributes its total bandwidth $B_u^{\max}$ across its DAs:
$\sum_{d \in \mathcal{D}_u} b_{u,d} \leq B_u^{\max}, \quad \forall u \in \mathcal{U}$, where $b_{u,d}$ is the bandwidth allocated to DA $d$ by UAV $u$. At level 2, within each DA, the allocated bandwidth $b_{u,d}$ is shared among users satisfying: $\sum_{e \in \mathcal{E}_d} b_e \leq b_{u,d}, \quad \forall d \in \mathcal{D}_u \label{eq:band_ue}$, where $\mathcal{E}_d$ denotes the set of users in DA $d$ and $b_e$ is the bandwidth allocated to user $e$ in a given scheduling slot.

This hierarchy naturally maps to a two-timescale operation framework for UAV-BSs, where they perform two allocation levels at different temporal granularities. At the \textit{large timescale} (in seconds), UAVs jointly decide: (i) 3D positioning, (ii) UE-to-UAV association and DA formation, and (iii) DA-level bandwidth allocation. At the \textit{small timescale} (in milliseconds), resource blocks (RBs) are dynamically assigned to UEs within each DA based on instantaneous channel states. This separation reflects the distinct dynamics: UAV mobility and aggregate demand evolve slowly, while per-user channel conditions fluctuate rapidly. \textit{This work only focuses on the large-timescale optimization problem; small-timescale scheduling employs Round-Robin for computational tractability.}

\vspace{-0.3em}
\subsection{Partial Observability}

Each UAV $u$ makes decisions based on locally observable information, including its own state (position, power, and resource status) and locally measurable network conditions (user demands, received interference from nearby UAVs). Although global state sharing is technically feasible, we adopt partial observability to preserve scalability. Full state exchange would incur substantial communication overhead, enlarge state-space dimensionality, and increase training complexity.


\vspace{-0.3em}
\section{Problem Formulation}


Our optimization problem involves the following decision variables for each UAV $u$ at the beginning of each large-timescale period: \textit{UAV's 3D position} $\mathbf{p}_u^t = (x_u^t, y_u^t, h_u^t)$, \textit{transmission power} $P_u^t$, and \textit{bandwidth allocation} $\mathbf{b}_u^t = [b_{u,d_1}^t, b_{u,d_2}^t, \ldots, b_{u,d_{|\mathcal{D}_u|}}^t]$. We maximize the weighted system performance presented in (\ref{eq:main-objective}):
\begin{align}
\max_{\{(\mathbf{p}_u^t, P_u^t, \mathbf{b}_u^t)\}} \;& 
\sum_{t=1}^{\infty} \big[ \alpha \Phi_{\text{QoS}}^t - \beta \Phi_{\text{Energy}}^t + \gamma \Phi_{\text{Fairness}}^t \big],
\label{eq:main-objective} \\
s.t. & \sum_{d \in \mathcal{D}_u^t} b_{u,d}^t \leq B_u^{\max}, \quad \forall u, t, \label{eq:bandwidth-constraint} \\
& \|\mathbf{p}_u^{t+1} - \mathbf{p}_u^t\| \leq v_{\max} \cdot T_L, \quad \forall u, t,\label{eq:mobility-constraint}\\
& (x_u^t, y_u^t, h_u^t) \in \mathcal{A}_{\text{service}}, \quad \forall u, t, \label{eq:area-constraint} \\
& P_{\min} \leq P_u^t \leq P_{\max}, \quad \forall u, t, \label{eq:power-constraint}
\end{align}
where $\alpha$, $\beta$, and $\gamma$, are weighting parameters that reflect the importance of objective components including \textit{QoS Satisfaction, Energy Efficiency}, and \textit{Fairness}. Constraint \eqref{eq:bandwidth-constraint} ensures each UAV respects its spectrum budget, \eqref{eq:mobility-constraint} limits UAV speed to $v_{\max}$ over large-timescale duration $T_L$, \eqref{eq:area-constraint} confines UAVs to the service area $\mathcal{A}_{\text{service}}$, and \eqref{eq:power-constraint} bounds transmit power.

\subsubsection{QoS Satisfaction Component} The QoS satisfaction component quantifies how well the achieved QoS including throughput, delay, and reliability of UEs meets their slice-specific requirements: $\Phi_{\text{QoS}}^{t} =\frac{\sum_{e \in \mathcal{E}}\omega_e\overline{S}_{e}^t}{\sum_{e \in \mathcal{E}} \omega_e}$,
where $\omega_e$ denotes the priority weight of slice type of UE $e$. The average satisfaction level  of UE $e$ at time $t$ is defined as:
\begin{equation}
\begin{split}
\overline{S}_{e}^t &= \omega^t_{e} \cdot \min\left(\frac{T_e^t}{T_e^{\min}}, 1\right) + \omega^r_e \cdot \min \left(\frac{R_e^t}{R_e^{\min}}, 1 \right)  \\ &+ \omega^d_s \cdot \min\left(1, \exp(\frac{D^{\max}_e-D^t_e}{D^{\max}_e})\right),
\end{split}
\end{equation}
where $\omega^t_{e}$, $\omega^d_{e}$, and $\omega^r_{e}$ represent the weighting coefficients that determine the relative importance of throughput, delay, and reliability for UE $e$, respectively, satisfying $\omega^t_{e} + \omega^d_{e} + \omega^r_{e} = 1$. $T_e^t$ and $T_e^{\min}$ denote the achieved and the minimum required throughput, respectively. $D_e^t$ and $D_e^{\max}$ represent the actual and maximum tolerable delay. Similarly, $R_e^t$ and $R_e^{\min}$ denote the achieved and the minimum required reliability of UE $e$, respectively. This formulation captures the overall QoS of each UE by jointly considering performance across multiple criteria. 

\subsubsection{Energy Penalty Component} 


This component penalizes excessive power usage and promotes during operation:
\begin{equation}
\Phi_{\text{Energy}}^{t} = \frac{1}{|\mathcal{U}|} \sum_{u \in \mathcal{U}} \frac{E^t_u}{E_u^{\max}},
\end{equation}
where $E^t_u$ is the energy consumption of UAV $u$ at time $t$, and $E_u^{\max}$ is its maximum possible consumption at each step. The energy consumption comprises: $E^t_u = P_u \cdot T_L + P_u^{\text{move}}\Delta \mathbf{p}_u$,
where $P_u^{\text{move}}$ is the propulsion power required for trajectory adjustment $\Delta \mathbf{p}_u$ over duration $T_L$ of the large timescale.

\subsubsection{Fairness Component} This component assesses the equality of resource distribution based on the throughput of each UE, modeled using the Jain’s fairness index as:
\begin{align}
\Phi_{\text{fair}}^{t} &=
\frac{\left( \sum_{e \in \mathcal{E}} x_e^{t} \right)^2}{|\mathcal{E}| \cdot \sum_{e \in \mathcal{E}} \left( x_e^{t} \right)^2},
\end{align}
where $x_e^{t}$ is the \textit{satisfaction/demand} ratio over the throughput of UE $e$ at time $t$, computed as $x_e^{t} = \min \left(\frac{T_e^t}{T^{\min}_e}, 1 \right)$. The resulting fairness index $\Phi_{\text{fair}}^{t}$ lies within the range $[0,1]$, where $\Phi_{\text{fair}}^{t} = 1$ indicates perfect equality in allocation across all UEs.


The formulated problem is computationally intractable for several reasons. The objective function couples continuous variables ($\mathbf{p}_u^t, P_u^t, \mathbf{b}_u^t$) through nonlinear functions (Shannon capacity, SINR in \eqref{eq:sinr}), creating a nonconvex landscape with multiple local optima. Inter-UAV interference introduces dependencies among agents, making the problem non-separable. The joint action space dimension scales with the number of UAVs and DAs. 
Traditional optimization methods such as successive convex approximation require iteratively solving multiple convex subproblems at each time step, with computational complexity that makes real-time execution infeasible. This motivates our learning-based approach, which can efficiently discover near-optimal policies through experience.

\vspace{-0.5em}
\section{Dec-POMDP Formulation for Spectrum Slicing in UAV-enabled Network}
We reformulate the optimization problem \eqref{eq:main-objective} as a DEC-POMDP \cite{bernstein2013complexitydecentralizedcontrolmarkov}. This formulation naturally captures three key characteristics of our system: (i) \textit{decentralized decision-making}: each UAV acts autonomously based on local information, (ii) \textit{partial observability}: UAVs cannot directly observe the complete system state due to communication constraints, and (iii) \textit{cooperative objectives}: all agents share the common goal of maximizing network-wide performance.

\subsection{Global State Space} The global state is defined as the joint observations of all UAV-BSs, encapsulating their spatial locations, transmission power levels, and bandwidth allocation profiles. Specifically, the local state of UAV $u$ is denoted as $\mathbf{\Omega}_u \stackrel{\text{def}}{=} \{ \mathbf{p}_u, P_u, \mathbf{D}_u \}$, where $\mathbf{p}_u$ represents the UAV’s position, $P_u^t$ is its current transmission power, and $\mathbf{D}_u$ denotes the demand-area allocation vector. Each element in $\mathbf{D}_u$ corresponds to a specific DA and encapsulates information including the number of active UEs, slice type, allocated bandwidth, and distance level. Accordingly, the system state space $\mathcal{S}$ is defined as the set of all UAV agents’ local states: $\mathcal{S} \triangleq \{(s)\} = \{ (\mathbf{\Omega}_u | \forall u \in \mathcal{U}) \}$.

\vspace{-0.5em}
\subsection{Local Observation Spaces} Each UAV agent $u$ receives a local observation that is available within its sensing range. The observation comprises two principal components: (i) \textit{UAV local state features} ($\mathbf{\Omega}_u$) captures the UAV’s intrinsic operational state, including its current position, transmission power, and bandwidth allocation status; (ii) \textit{Environmental context} ($\mathbf{I}_u$) maintains spatial awareness without a complete global state, each UAV additionally observes its surrounding environment, including the distribution of active UEs within its area of interest, and the relative density of neighboring UAVs. Both quantities are estimated along the four cardinal directions (North, East, South, and West). These coarse-grained features provide contextual awareness for inter-agent coordination. Accordingly, the local observation space of UAV agent $u$ is defined as: 
$\mathcal{O}_u \triangleq \{ (\mathbf{\Omega}_u, \mathbf{I}_u) \}.$

\vspace{-0.5em}
\subsection{Action Space} Based on its local observation, each UAV agent $u$ selects an action $a_u$ consisting of three continuous control variables: the positional adjustment $\Delta \mathbf{p}_u$, the transmission power $P_u$, and the spectrum allocation vector $\mathbf{b}_u$. These control variables respectively correspond to mobility control, power adaptation, and resource allocation. Hence, the action space of UAV agent $u$ is defined as $\mathcal{A}_u \triangleq \{ \Delta \mathbf{p}_u, P_u, \mathbf{b}_u \}$.

\vspace{-0.5em}
\subsection{Reward Function} All UAVs share a global reward signal consistent with the optimization objective in Eq.~\eqref{eq:main-objective} as follows:
\begin{align}
r(s_t, \mathbf{a}_t) = \alpha \cdot \Phi_{\text{QoS}}^{t}
- \beta \cdot \Phi_{\text{Energy}}^{t}
+ \gamma \cdot \Phi_{\text{Fairness}}^{t}.
\end{align}
Since all agents observe the same reward signal, the problem exhibits a \textit{cooperation} where agents must coordinate to maximize collective performance rather than individual gains.

\vspace{-0.5em}
\subsection{Policy and Objective}

Each agent $u$ learns a deterministic policy $\pi_u: \mathcal{O}_u \rightarrow \mathcal{A}_u$ that maps local observations to continuous actions. The joint policy is $\boldsymbol{\pi} = (\pi_1, \ldots, \pi_U)$. The goal is to find the optimal joint policy $\boldsymbol{\pi}^*$ that maximizes the expected cumulative discounted reward:
\begin{equation}
\boldsymbol{\pi}^* = \arg\max_{\boldsymbol{\pi}} \; \mathbb{E}_{\boldsymbol{\pi}} \left[ \sum_{t=0}^{\infty} \gamma^t r(s_t, a_t) \; \bigg| \; a_t \sim \boldsymbol{\pi}(o_t) \right],
\label{eq:policy-objective}
\end{equation}
where the instantaneous reward $r(s_t, a_t)$ reflects the network performance achieved at time step $t$ under the joint action of all UAVs following policy $\boldsymbol{\pi}$ in the system state $s_t$. $\gamma \in [0,1)$ is the discount factor. In the subsequent section, we introduce the multi-agent reinforcement learning framework employed to solve this DEC-POMDP optimization problem.

\section{MADDPG Framework with Shared Critic}

To solve the Dec-POMDP formulated in Section~IV, we employ a Multi-Agent Deep Deterministic Policy Gradient (MADDPG) framework \cite{lowe2017maddpg} adapted to the CTDE paradigm. CTDE leverages global information during training to stabilize learning while maintaining decentralized execution for scalability. Our framework introduces a innovations that is the \textbf{shared critic} architecture that exploits the cooperative nature of UAV agents.

\vspace{-0.5em}
\subsection{Actor Networks} Each UAV agent $u \in \mathcal{U}$ employs an individual actor network $\pi_u(\cdot; \theta_u)$ that maps its local observation $o_u^t \in \mathcal{O}_u$ to a continuous action $a_u^t = \pi_u(o_u^t; \theta_u) + \epsilon_t$,
where $\epsilon_t$ denotes the exploration noise. Gaussian noise is applied to the movement and power actions, while a Dirichlet distribution perturbs the bandwidth allocation to preserve valid probabilistic constraints. The exploration variance decays over time as: $\sigma_t = \sigma_0 \cdot \rho^t, \quad (0 < \rho < 1$), allowing the agents to gradually shift from exploration to exploitation.

\vspace{-0.5em}
\subsection{Shared Critic Network} The centralized critic $Q(s, \mathbf{a}; \phi)$ is parameterized by $\phi$ and estimates the joint action-value function given the state $s$ and joint action $\mathbf{a} = [a_1, \ldots, a_U]$ as $v_t = Q(s^t, \mathbf{a}^t; \phi)$. The critic integrates all agents' observation-action embeddings through a multi-head self-attention module, enabling adaptive weighting of inter-agent dependencies. This design enhances the critic's ability to capture cooperative relationships among UAVs and improves stability during policy updates.

\vspace{-0.5em}
\subsection{Target Actor and Critic Networks} To stabilize training, each actor and critic has a corresponding \textit{target network}, denoted as $\pi'_u(\cdot; \theta'_u)$ and $Q'(s, \mathbf{a}; \phi')$, respectively. These networks are delayed copies of the main networks and are used to compute stable target values during updates. By updating them slowly, the learning process avoids oscillations caused by rapidly changing policy and value estimates: $\phi' \leftarrow \tau \phi + (1 - \tau)\phi'$; $\theta'_u \leftarrow \tau \theta_u + (1 - \tau)\theta'_u, \forall u \in \mathcal{U}$, where $\tau \in (0,1)$ is the soft update rate.

\vspace{-0.5em}
\subsection{Actor and Critic Loss} 
Each actor $u$ is optimized to maximize the expected return estimated by the centralized critic. The gradient of each actor $u$ policy is obtained by minimizing:
\begin{equation}
\mathcal{L}_{\pi_u}(\theta_u) = -\mathbb{E}_{s \sim buffer}
\left[ Q(s, a_1, \ldots, a_u, \ldots, a_U; \phi) \right],
\label{eq:actor_loss}
\end{equation}
where $a_u = \pi_u(o_u; \theta_u)$ for the active agent and $a_v$ is got from experience buffer for all $v \neq u$ to prevent gradient interference between actors. A experience \textit{buffer} stores the tuple $(\mathbf{o}^t, \mathbf{a}^t, r^t, \mathbf{o}^{t+1})$ at each step. Mini-batches are uniformly sampled during training to decorrelate consecutive transitions and improve sample efficiency. The critic parameters $\phi$ are updated by minimizing the following loss:
\begin{equation}
\mathcal{L}_Q(\phi) = \mathbb{E}_{(\mathbf{s}, \mathbf{a}, r, \mathbf{s}') \sim buffer} \left[ \left( y - Q(s, \mathbf{a}; \phi) \right)^2 \right],
\label{eq:critic_loss}
\end{equation}
where the temporal difference target is computed using the target networks as:  $y = r + \gamma Q'(s', \mathbf{a}'; \phi')$, with $\mathbf{a}' = \{\pi'_u(o'_u; \theta'_u)\}_{u=1}^U$ representing the target actions.

\subsection{Training Algorithm}

The training procedure of the proposed MADDPG-based framework is summarized in \textbf{Algorithm~\ref{alg:ctde}}.

\begin{algorithm}[H]
\caption{MADDPG Training}
\begin{algorithmic}[1]
\label{alg:ctde}
\STATE Initialize actor networks $\{\pi_u\}_{u=1}^U$, centralized critic $Q$, corresponding target networks, and relay buffer;
\FOR{$t = 1$ to $T$}
    \FOR{each agent $u \in \mathcal{U}$}
        \STATE Select action $a_u^t = \pi_u(o_u^t; \theta_u) + \epsilon_t$;
    \ENDFOR
    \STATE Execute joint action $\mathbf{a}^t$ and observe $(r^t, \{o_u^{t+1}\}_{u=1}^U)$;
    \STATE Store $(\mathbf{o}^t, \mathbf{a}^t, r^t, \mathbf{o}^{t+1})$ in buffer;
    \IF{$|\text{buffer}| >$ batch\_size}
        \STATE Update critic as in (\ref{eq:critic_loss});\
        \FOR{each agent $u \in \mathcal{U}$}
            \STATE Update actor $u$ as in (\ref{eq:actor_loss});
        \ENDFOR
        \STATE Perform soft updates of target networks;
    \ENDIF
\ENDFOR
\end{algorithmic}
\end{algorithm}

\section{Numerical Results}

\subsection{Simulation Settings}

In the simulation setup, a service area of \(2000 \times 2000\,\mathrm{m}\) is considered, where the number of UEs fluctuates from 180 to 300. Three UAV-BSs are deployed, each operating within the bounds \([0,2000]\)~m on the \(x\)-axis, \([0,2000]\)~m on the \(y\)-axis, and \([100,400]\)~m on the \(z\)-axis. The DAs in each slice are classified by distance: \textit{Near} (0-300~m), \textit{Medium} (300-600~m), and \textit{Far} ($>$600~m). The noise power density is set to \(10^{-13}\,\mathrm{W}\) and a carrier frequency of 3.5~GHz. Each UAV-BS is allocated a total bandwidth of 600~MHz (1667 RBs each). UAVs transmit signal with power in the range of 0-10 W, and beam angle of $60^\circ$.

Each UAV agent employs an actor network that processes an 80-dimensional observation vector through specialized encoders: (i) a UAV state encoder for position and transmit power, (ii) a DA information encoder with two hidden layers capturing queue and delay features, and (iii) a surrounding environment encoder. A multi-head self-attention mechanism is applied to the DA representation to enhance bandwidth allocation adaptability. The centralized critic evaluates joint state-action pairs by concatenating all agents’ observations and actions, followed by a six-layer feedforward network.

The MADDPG framework employs actor and critic learning rates of $\eta_\pi = 10^{-4}$ and $\eta_Q = 10^{-3}$, respectively, with discount factor $\gamma = 0.99$ and soft target update rate $\tau = 0.005$. Training utilizes a replay buffer of 300,000 transitions with mini-batch size 256. Gaussian exploration noise is applied to continuous actions (position, power) with initial variance $\sigma_0 = 1.0$, decaying to $\sigma_{\min} = 0.005$ via decay rate $\rho = 0.981$. Dirichlet noise perturbs bandwidth allocation to maintain simplex constraints. The network is trained until the exploration noise hit $\sigma_{\min}$ and then the exploration noise is reset to 0.6 for re-exploration. We benchmark the proposed method against three strategies: 
\begin{enumerate}
    \item \textbf{Random}: Uniformly samples actions from the valid action space.
    \item \textbf{Coverage-Greedy}: Positions UAVs at user centroids with altitude $h_u = 1.1 d_{\max} / \tan(60^\circ)$ to ensure beam coverage, where $d_{\max}$ is the horizontal distance to furthest assigned UE. Power scales with altitude; bandwidth favors far DAs over near DAs.
    \item \textbf{QoS-Greedy}: Adapts altitude to slices (URLLC: 30\% height, mMTC: 70\% height) using weighted average. Positions toward the weighted centroid of UEs (URLLC: 3×, eMBB: 2×, mMTC: 1×).
\end{enumerate}

\vspace{-0.5em}
\subsection{Simulation Results}

\vspace{-0.5em}
\begin{figure}[ht]
    \centering
    \subfloat[]{%
        \includegraphics[width=0.48\linewidth]{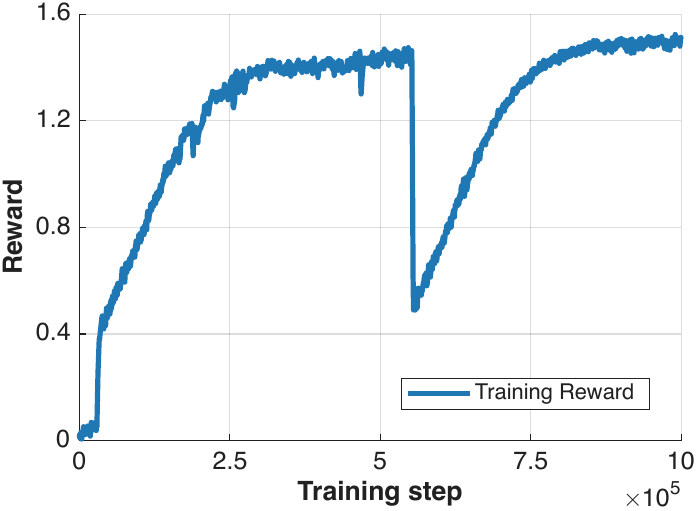}
        \label{fig:training-reward}
    }
    \hfill
    \subfloat[]{%
        \includegraphics[width=0.49\linewidth]{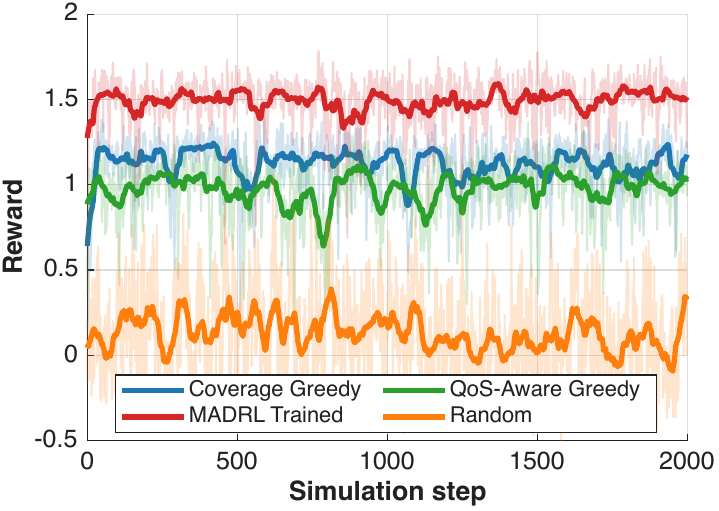}
        \label{fig:evaluation-reward}
    }
    \caption{Evaluation of the MADDPG framework: (a) Training reward in 1M steps and (b) Reward comparison throughout 2000 simulation steps.}
    \label{fig:reward}
\end{figure}

\subsubsection{Overall Reward} In our simulation, we set the weights of 2.0, 0.8, and 0.5 for QoS reward, energy penalty, and fairness reward, respectively. As illustrated in Fig.\ref{fig:training-reward}, the training reward demonstrates a stable upward trajectory, indicating that the proposed MADDPG framework successfully converges to an optimal policy. Throughout 2000 steps of simulation, our method consistently outperforms all baselines in terms of overall reward (Fig.\ref{fig:evaluation-reward}). The performance of our method is approximately 33\% better compared to greedy methods. This confirms the agents’ ability to learn effective strategies that balance multiple objectives, including QoS satisfaction, energy efficiency, and fairness through experience rather than relying on pre-defined heuristics.

\subsubsection{QoS Reward}
\begin{figure*}[ht]
    \centering
    \subfloat[]{%
        \includegraphics[width=0.325\linewidth]{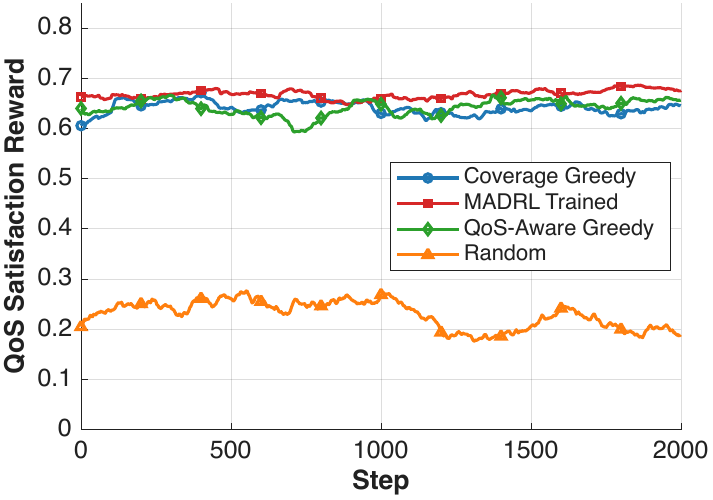}
        \label{fig:qos-reward}
    }
    \hfill
    \subfloat[]{%
        \includegraphics[width=0.325\linewidth]{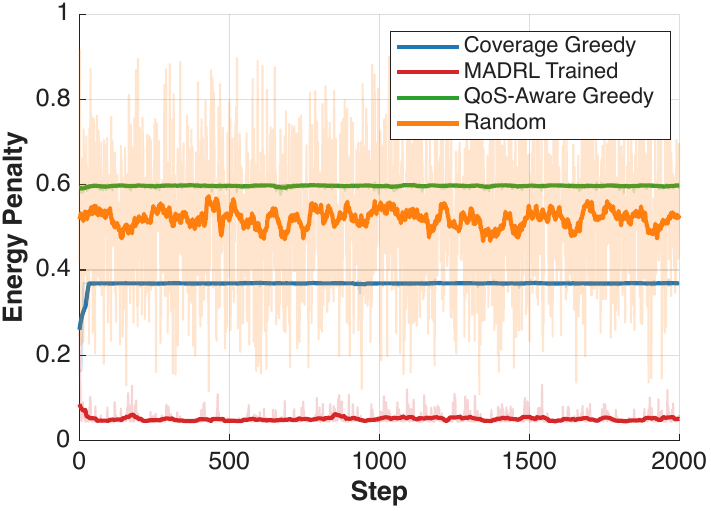}
        \label{fig:energy-penalty}
    }
    \hfill
    \subfloat[]{%
        \includegraphics[width=0.325\linewidth]{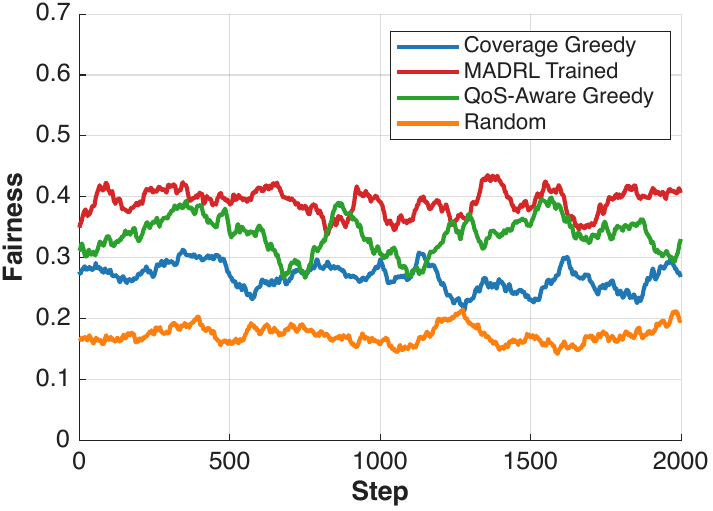}
        \label{fig:fairness-reward}
    }
    \caption{Breakdown of the total reward: (a) QoS reward, (b) Energy Penalty, and (c) Fairness reward.}
    \label{fig:reward-components}
\end{figure*}

In terms of QoS performance (Fig.\ref{fig:qos-reward}), the MADDPG approach achieves a slightly superior level of service satisfaction (approximately 66\%) compared to the greedy baselines (4\% better on average), while significantly outperforming the random policy. Unlike heuristic methods such as the QoS-aware or Coverage Greedy approaches, which depend on fixed thresholds for RBs allocation, the MADDPG agents dynamically learn to allocate bandwidth based on real-time observations. For instance, given similar data demands, the framework learns to prioritize URLLC slices at certain DAs with higher reliability requirements, ensuring adaptive and context-aware resource distribution. Furthermore, through cooperation, agents learn to coordinate trajectories to mitigate interference due to overlapping regions among UAV-BSs.

\subsubsection{Energy Penalty}

Energy consumption results (Fig.\ref{fig:energy-penalty}) reveal a substantial improvement of MADDPG compared to all baselines. While higher transmission power can enhance QoS, it also leads to rapid energy depletion. The proposed framework effectively learns this trade-off, adapting to sustain acceptable service quality while conserving energy. The observed reduction in energy consumption demonstrates that the agents have learned to perform efficient trajectory control and power management with up to $8\times$ less energy consumption compared to other baselines. This yields a substantial reduction in energy consumption while preserving QoS.

\subsubsection{Fairness Reward}
Fairness evaluation (Fig.\ref{fig:fairness-reward}) highlights the superior adaptability of our approach. In radio access networks. MADDPG achieves higher fairness indices compared to greedy and random strategies. MADDPG has an average of 16\% better fairness compared to the QoS-aware. This is attributed to the agents’ ability to cooperatively reason about local and global network states, resulting in balanced RB allocation across UEs with diverse QoS requirements. Consequently, MADDPG ensures more equitable service provisioning and improved user experience.

\section{Conclusion}
This paper has presented a hierarchical network slicing framework for UAV-satellite integrated networks, formulated as a Dec-POMDP that jointly optimizes trajectory, power, and spectrum allocation. We have developed a shared-critic MADDPG solution that enables UAV agents to learn cooperative policies while maintaining decentralized execution. Simulation results have demonstrated that our approach achieves a 33\% improvement in cumulative reward over baseline methods while reducing energy consumption by up to $8\times$ and improving fairness by 16\% on average. These results have validated that learned policies can effectively balance QoS satisfaction, energy efficiency, and fairness through multi-agent cooperation. Future work includes extending the framework to incorporate Rician fading with time-varying K-factors and Doppler effects, investigating heterogeneous UAV fleets with mixed altitude and capability profiles, and conducting experimental validation on hardware-in-the-loop testbeds.

\bibliographystyle{IEEEtran}
\bibliography{IEEEfull}

\end{document}